\documentclass[fleqn,usenatbib]{mnras}

\usepackage{newtxtext,newtxmath}
\usepackage[T1]{fontenc}
\usepackage{ae,aecompl}

\usepackage{graphicx}	% Including figure files
\usepackage{amsmath}	% Advanced maths commands
\usepackage{wasysym}
\usepackage{CJKutf8}
\newcommand{\degree}{^\circ}
\title[Earth--Mars Belt]{
Four Billion Year Stability of the Earth--Mars Belt
}

\author[Huang \& Gladman]{
Yukun Huang (\begin{CJK*}{UTF8}{gbsn}黄宇坤\end{CJK*})$^{1}$ and
Brett Gladman$^{1}$
\\
$^{1}$Department of Physics and Astronomy, University of British Columbia, 6224 Agricultural Road, Vancouver, BC V6T 1Z1, Canada
}

\date{Accepted XXX. Received July xx/2020; in original form ZZZ}

\pubyear{2020}

\begin{document}
\label{firstpage}
\pagerange{\pageref{firstpage}--\pageref{lastpage}}
\maketitle

% Abstract of the paper2
\begin{abstract}
	Previous work has demonstrated orbital stability for 100 Myr of initially near-circular and
	coplanar small bodies in a region termed the `Earth--Mars belt' from  1.08~au $< a <$ 1.28~au.
	Via numerical integration of 3000 particles, we studied orbits from 1.04--1.30~au for the age of
	the Solar system.
	We show that on this time scale, except for a few locations where mean-motion resonances
	with Earth
	affect stability, only
	a narrower `Earth--Mars belt' covering  $a\sim(1.09 ,1.17 )$ au, $e<0.04$, and $I<1\degree$
	has over half of the initial orbits survive for 4.5~Gyr.
	In addition to mean-motion resonances, we are able to see how the
	$\nu_3$, $\nu_4$, and $\nu_6$
	secular resonances contribute to long-term instability in
	the outer (1.17--1.30~au) region on Gyr time scales.
	We show that all of the (rather small) near-Earth objects
	(NEOs) in or close to the Earth--Mars belt appear to be
	consistent with recently arrived transient objects by
	comparing to a NEO steady-state model.
	Given the $<200$ m scale of these NEOs, we estimated the Yarkovsky drift rates in semimajor axis, and use these to estimate that primordial asteroids with a diameter of 100 km or larger in the Earth-Mars belt would likely survive.
	We conclude that only a few 100-km sized asteroids could have been present in the belt's region at the end of the terrestrial
	planet formation.
\end{abstract}


%%%%%%%%%%%%%%%%%%%%%%%%%%%%%%%%%%%%%%%%%%%%%%%%%%

%%%%%%%%%%%%%%%%% BODY OF PAPER %%%%%%z%%%%%%%%%%%%

\section{Introduction}\label{sec:intro}

%WORK IN BELOW we can possibly predict where to find more

The Solar System's small bodies are often regarded as more primitive relics dating back to early stages of the Solar System.
Their current orbital distribution sheds light on planetesimal/planet formation, if the objects are still located in their primordial location.
It is thus necessary to diagnose which dynamical niches have a
large-enough phase-space volume to make survival of primordial populations
residing at those locations $\sim$4 Gyr ago feasible.
Such stability is clearly necessary for primordial retention, but there may be regions of such stability which are not currently populated; such empty niches then demand that early dynamical processes emptied them of large bodies.

The two major minor-body belts in the Solar System are the main asteroid belt
and the Kuiper belt.
The main part of the asteroid belt (with semimajor axes $a\simeq$ 2--3~au)
is sufficiently far from the neighbouring planets that
eccentricities must reach $e$ = 0.2--0.4 before
planet crossing with Mars or Jupiter occurs.
% At the end of planet formation (when the planets reached their current orbits), (revision)
After the planets reached their current orbits,
many asteroids were still present in this region and subsequent dynamical erosion 
%has only eliminated a small fraction of them 
has not reduced the population by more than a factor of a few 
\citep{2015aste.book..493M},
although mass is moving steadily to smaller mass bins by collisional activity.

%\textcolor{red}{INTRO TO THE MAIN BELT}
%Most of the minor bodies orbit the sun in a stable region between the orbits of Mars and Jupiter (also known as the main asteroid belt). \textcolor{red}{WHAT BOUNDS ASTEROID BELT?  Mars and Jupiter crossing requires e 0.2-0.4 before
%	planet crossing happens.
%	At end of planet formation, many bodies were present in this region and subsequent collisional and dynamical erosion has removed some.}

%\textcolor{red}{INTRO TO THE KUIPER BELT}
%In the outer Solar System, beyond the orbit of Neptune, the Kuiper belt is believed to be a reservoir hosting a large amount of icy bodies, whose total mass exceeds that of the asteroid belt by two orders of magnitude. The last decades has seen a rapid growth in the number of discovered Kuiper belt objects (KBOs) or trans-Neptunian objects (TNOs) \citep{Bannister:2018ha}, and the populations of the resonant TNOs are crucial to the understanding of the outward migration of Neptune during the late stage of the giant planet formation \citep{1995Icar..118..132M,Gladman:2012ed}.

The large icy-body reservoir of the Kuiper belt is bounded on the inside by Neptune's ability to
rapidly remove objects with perihelia $q$ below about 35 au (except for some resonant objects
who can even have $q<30$~au due to the dynamical protection provided by the resonance
\citep{2012AJ....144...23G}).
The outer edge of the heavily populated belt is {\it not} set by a planet crossing limit, but
rather by poorly understood cosmogonic processes.
Low-$e$ and low-$i$ orbits are stable to great distances.

Although they are not `belts' covering a large range of semimajor axes,
the jovian and neptunian Trojans provide two more abundant populations
of small bodies that are stable on Solar System time scales.
It should be noted that the Neptune's Trojans consist of a mix of objects
which are primordial (with stability time scale $> \sim4$ Gyr) and a set which
is most likely recent temporary captures out of the scattering/Centaur population \citep{2016AJ....152..111A}.
That is, some unstable small bodies are able to temporarily `stick' to
resonant niches and provide a non-primordial component.
Estimated numbers (factor of 2) of absolute magnitude $H<9$ objects in the asteroid main belt,
jovian Trojan, neptunian Trojan, and main Kuiper belt are
300, 30, 100, and 100,000, respectively \citep{2016AJ....152..111A, 2011AJ....142..131P}.

In addition to these stable populations, the planet-crossing near-Earth Object (NEO)
and Centaur populations have dynamical lifetimes much shorter than the age of
the Solar System, and thus must be resupplied from more stable structures.
The NEOs are predominantly replenished by leakage from the main asteroid belt,
which provides the known orbital and absolute magnitude distribution
\citep{2002Icar..156..399B, Greenstreet:2012ee, Granvik:2018jc},
while the Centaurs are likely dominantly supplied from the trans-neptunian scattering disk \citep{2008ApJ...687..714V}.

\subsection{Hypothetical long-lived regions}

In addition to the regions which are dynamically stable for 4.5 Gyr and are known
to host small-body populations, there are a few dynamical niches that  numerical studies have shown to be stable for the age of the Solar
System but which have no known members with demonstrated 4 Gyr stability.

The first such established case is for the
`Uranus--Neptune' belt near 26~au.
\citet{Gladman:1990th} demonstrated some stable orbits for their 22.5~Myr
maximum integration time, which
\citet{Holman:1993iw}
pushed to 800 Myr and demonstrated continued survival for this time.
\citet{Anonymous:CaEofV_6} extended the integration time to 4.5 Gyr, and found 0.3\% of initially near-circular and near-planar orbits between the orbits of Uranus and Neptune (24-27~au) survive for the age of the Solar System.
There are still no known bodies discovered in the stable region (despite
great sensitivity to them in trans-neptunian surveys), and
\citet{Anonymous:CaEofV_6} and
\citet{1998Icar..135..408B}
argue that by the end of giant planet formation, it is likely that no bodies would remain at low $e$ and $i$ in this region.

Based on numerical integrations of hypothetical Earth Trojan asteroids,
\citet{2000MNRAS.319...63T}
demonstrated 50 Myr orbital stability of these objects and speculated by extrapolation
of the decay rate that some orbits could survive for the age of the Solar
system.
Earth co-orbitals in horseshoe orbits (that is, not librating around a single Lagrange point like Trojans, but instead encompassing 
three points)  have even longer stability time ($\sim$Gyr) than Earth Trojans have \citep{2012MNRAS.426.3051C}. It may thus be more likely to find a primordial Earth  horseshoe \citep{2019A&A...622A..97Z}, but all {\it known} Earth  co-orbitals (of any type) have much shorter dynamical lifespans than the Solar System's age (see references listed in \citet{Greenstreet:2020jx}).
Recently, \citet{2019A&A...622A..97Z}
concluded that the dynamical erosion over 4~Gyr, especially when accounting for
Yarkovsky drift, would eliminate a population of sub-km primordial
Earth co-orbitals surviving to the present day, but that km-sized could
survive.
Direct observational searches by \citet{2018LPI....49.1149C} and \citet{Markwardt:2020ec}
provided no Trojan detections down to sizes of a few hundred meters,
leading to the conclusion that it is unlikely that any Earth Trojans
still exist.
Thus, simply demonstrating that the dynamics permit a portion of phase space
to be stable does not imply that a population currently exists; again, the perturbations
inherent in planet formation likely resulted in no large stable Earth co-orbitals being present at the end of planet formation.
The only temporary Earth Trojan 2011 TK7 \citep{2011Natur.475..481C} is dynamically unstable and not consistent with a primordial origin.
The known co-orbitals of the Earth are best explained as temporarily trapped
near-Earth asteroids \citep{MORAIS20021}.

Inside Mercury's orbit, the hypothesized population of small bodies, known as the Vulcanoids, was first proposed by \citet{1978Icar...35...99W}. \citet{Evans:wa} numerically studied the stability of the intra-mercurial region and found that the dynamical niche where Vulcanoids may exist is from 0.09~au to 0.21~au. However, accounting for Vulcanoid evolution under the Yarkovsky thermal force shows that objects with the diameter $<$1~km would be removed over the age of the Solar System \citep{2000Icar..148..147V}.  \citet{2020AAS...23527701C} calculated that even 100~km sized Vulcanoids could
rotationally fission in less than the Solar System's age.
A recent search for Vulcanoids with NASA's STEREO spacecraft \citep{2011epsc.conf..250S} returned no detection, and thus the existence of Vulcanoids larger than 5.7 km in diameter was ruled out (3$\sigma$ upper limit).

\subsection{An Earth--Mars Belt?}

\citet{Evans:wa} mapped stable orbits in both the Vulcanoid region and
a hypothetical belt between Earth and Mars (1.08 au $< a < $ 1.28 au).
They investigated the structure of those two belts in greater detail, exploring the role of mean-motion resonances with terrestrial planets, as well as the $\nu_6$ and $\nu_{16}$ secular resonances,
in sculpting the belts \citep{2002MNRAS.333L...1E}.

The most obvious limitation of the \citet{Evans:wa, 2002MNRAS.333L...1E} study is the short integration timescale. Their simulations only ran for 100 Myr (2\% of 4.5 Gyr), which posed an interesting question: Could the hypothetical Earth--Mars belt survive for the age of the solar system? To our knowledge, no study has further investigated the orbital stability of this inconspicuous region.
If the answer to this question is yes, one would immediately want to determine if
there is still a primordial asteroid alive in the Earth--Mars belt?
A sample return from such a primordial planetesimal would give unique insights into planetesimal
accretion in the inner Solar System.

In contrast, if no primordial asteroid is found in this stable region (despite the fact that the corresponding area of sky has been extensively and consistently covered by various NEO and asteroid surveys), this implies that
the belt is sparsely populated at the end of terrestrial planet formation.
The survival fraction from the dynamical simulations then provides a required upper limit on the
population of large objects that planet-formation simulations
can leave in the region once they form the terrestrial planets on their
current orbits.

% The paper is structured as follows: In Section \ref{sec:numerical}, we present our numerical simulation results and carry out detailed analysis on the dynamical stability and the structure of the Earth--Mars belt. In Section \ref{sec:nea}, we calculate the dynamical lifetime of NEOs that are currently inside or surrounding the Earth--Mars belt, with the aim of finding any existent primordial asteroid in the region. Section \ref{sec:discussions} contains a discussion of the Yarkovsky effect and the cosmogonic implications of our results. Section \ref{sec:conclusions} states our conclusions.

% -------------------- section ----------------------------

\section{Numerical Results and Analysis}\label{sec:numerical}

\begin{figure*}
	\includegraphics[width=\textwidth]{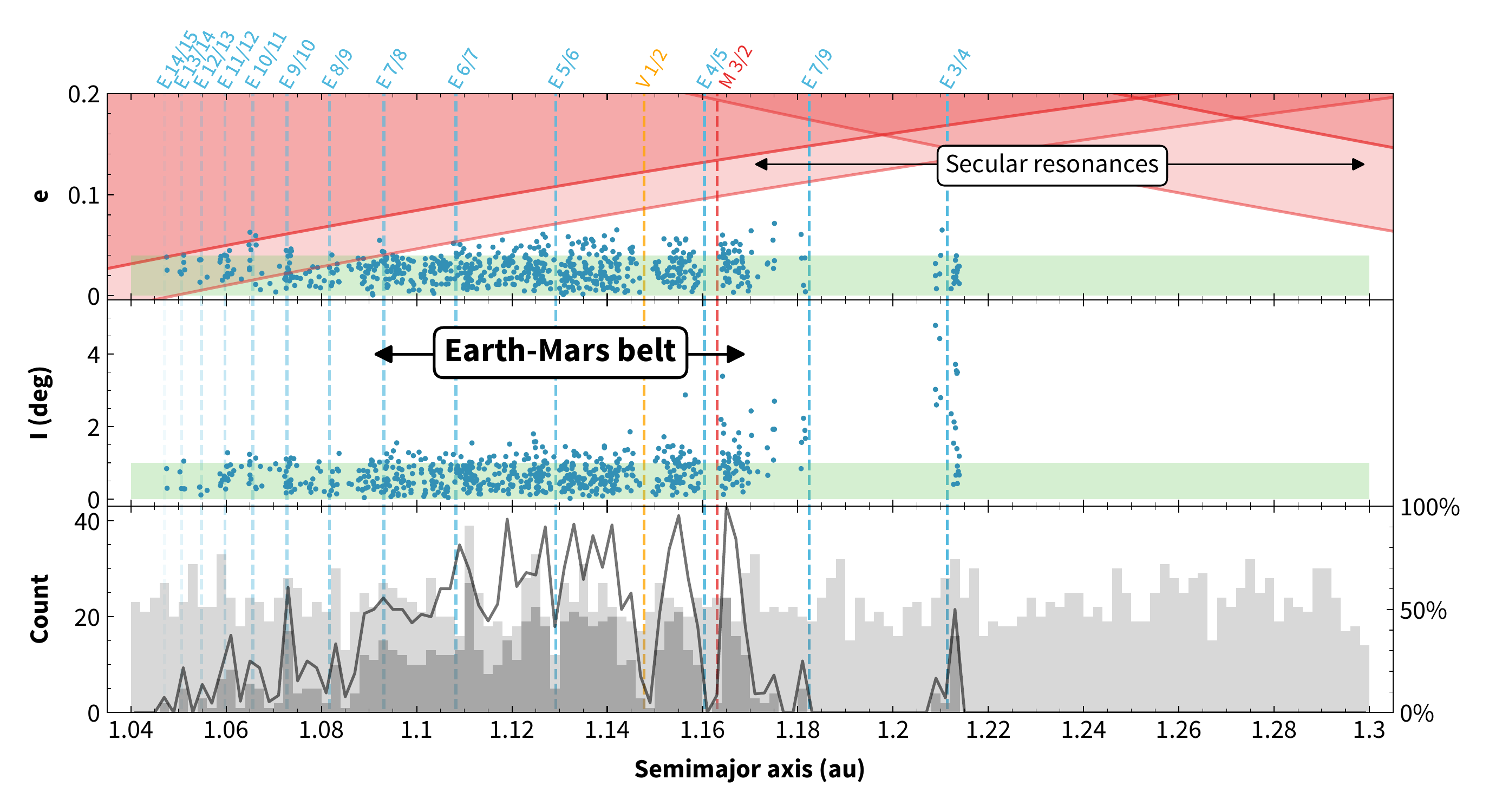}
	\caption{The vertical dashed lines mark all first-order resonances with Earth (blue) and Venus (orange), as well as the 3:2 with Mars (red) and
	the second-order 7:9 resonance with Earth.
	{\bf Top two panels:} $(a,e)$ and $(a,I)$ distributions of all surviving test particles at the end of the 4.5 Gyr integration time. The green rectangles represent the ranges of initial orbital elements, within which the initial $(a,e,I)$ were randomly generated. The top left and the top right corners of the first panel are covered by two red bands, showing the orbital intersecting bands (see text) related to Earth's aphelion and martian perihelion.
	The $a$ range over which secular resonances result in broad instability is indicated.
	The span of Earth-Mars belt, that we define in the text,
	is from 1.09 au to 1.17 au and is shown in the middle panel.
	{\bf Bottom panel:} The semimajor axis histogram of all test particles at the start (light grey bins) and surviving ones at the end (dark grey bins). The initial to final ratio for each bin is shown with the grey lines.}
	\label{fig:alltp}
\end{figure*}

We performed a detailed 4.5 Gyr numerical simulation, encompassing
the Earth-Mars belt region using the $\texttt{rmvs4}$ package of SWIFT \citep{1994Icar..108...18L}.
The initial orbital elements of 3000 test particles were randomly generated in near-circular and near-coplanar orbits, drawn at random between $a\sim(1.04, 1.3)$~au, $e\sim(0, 0.04)$,
J2000 ecliptic inclination
$I\sim(0,1\degree)$, and with random phase angles. The simulation were run for 4.5 Gyr with a time step of 0.01 yr.
The gravitational effects of 7 planets, excluding Mercury, were taken into account.
Another parallel simulation, in which Mercury was included, was run for 1 Gyr to verify our results and we found no statistically significant difference between the two simulations;
Mercury thus has no significant influence on the stability of the fictitious Earth--Mars belt and was excluded in our main 4.5~Gyr simulation.
Test particles were removed from the simulation upon approaching within 0.01~au of a planet (for Earth,
this corresponds to its Hill sphere).
Although an inner and outer bound (of the solar radius and 5.2~au, respectively) existed, no particles reached this
state before a planetary encounter.

Over the entire initial semimajor axis range,
679/3000 (22.6\%) of the test particles survive for the age of the Solar System;
%, whose results
the final states of the surviving particles
are shown in Figure \ref{fig:alltp}.
The $(a, e)$ and $(a, I)$ distributions
%for particles remaining at the end of the integration
are plotted in the first two panels; the third panel gives histograms of
initial and final numbers and their ratio for each semimajor bin.
We inspected the integrations and found that semimajor axis changes $\Delta a$ for surviving particles are very rarely exceed the bin width of 0.002~au,
and thus the ratio is very close to the survival fraction for the initial
bin.  (The number per initial bin is non-uniform because the initial $a$ was chosen randomly).
Based on these results, we chose to divide the semimajor range into three different regions: A strongly unstable region ($a$ > 1.17 au), the
resonance region ($a$ < 1.09 au), and the main Earth-Mars belt (1.09 au < $a$ < 1.17 au). The final/initial fractions are 36/1476 (2.4\%) for the unstable region, 102/587 (17.4\%) for the resonance region, and 541/916 (59.1\%) for the Earth-Mars belt.
We note that nearly 80\% of the survivors reside in the belt.

As pointed out by \citet{Evans:wa}, the Earth--Mars belt is sculpted by various mean-motion and secular resonances.
We carry out the analysis on the role of these resonances in greater detail.
A $k^\prime:k$ mean motion resonance with a planet occurs when $kn - k^\prime n^\prime \approx 0$, where $n$ and $n^\prime$ are the mean motion frequencies of the asteroid and the planet, respectively. $|k^\prime-k|$ is defined as the order of the resonance, and the exact resonant location can be obtained by applying
$a = a^\prime \; (k/k^\prime)^{\frac{2}{3}}$.
First order resonances with the three terrestrial planets, as well as a second order resonance with Earth (7:9), are plotted and colour-coded in Figure \ref{fig:alltp}.

Although the Earth's semimajor axis is almost unchanging during the whole integration, secular perturbations provided by other planets constantly alter its eccentricity.
We filtered Earth's orbital eccentricity history, and found that Earth's $e$ spends 90\% of the time fluctuating between 0.007 and 0.049. The two red solid curves on the top left panel indicates $a_\oplus (1+e_\oplus) = a(1-e)$, where $e_\oplus$ denotes the 90\% range of Earth's eccentricity.
Likewise, the Martian intersecting band is $a_{\mars} (1-e_{\mars}) = a(1+e)$, where the 90\% range of its eccentricity is (0.018, 0.089).

In the resonance region ($a$ < 1.09 au), the integration shows that asteroid orbits intersect with Earth's due to proximity, and thus require protection.
For a circular orbit for Earth, we first note that for initial $a< 1 + 2.4(m_\oplus/M_\odot)^{1/3} = 1.035$~au, test particles that begin with $e\simeq0$ will be in the `crossing zone' and have Earth encounters
\citep{Gladman:1993fg}.
First-order mean-motion resonance overlap generates dynamical chaos (but not
necessarily planet-crossing) out to $1 + 1.5(m_\oplus/M_\odot)^{2/7} = 1.040$~au \citep{1980AJ.....85.1122W, 1989Icar...82..402D}.
Our calculations show that out to 1.09~au, the inner border of the Earth--Mars belt is perpetually swept by the Earth-intersecting band as Earth's eccentricity varies, and only particles trapped in the (now non-overlapping) mean-motion resonances survive.
These resonant particles are phase protected from close encounters with the planet, allowing long-term stability near the resonant semimajor axes.
As the histogram shows,
the resonant locations (vertical lines) with $a<$1.09~au
coincide with clusters of stable particles.
This kind of resonance protection (with particle perihelia smaller than a planet's aphelion) are commonly found in the resonant transneptunian population, which is a major component of the Kuiper Belt in the outer Solar System
\citep[{\it e.g.},][]{2018ApJS..236...18B}.

%BGBG

\begin{figure}
	\includegraphics[width=\columnwidth]{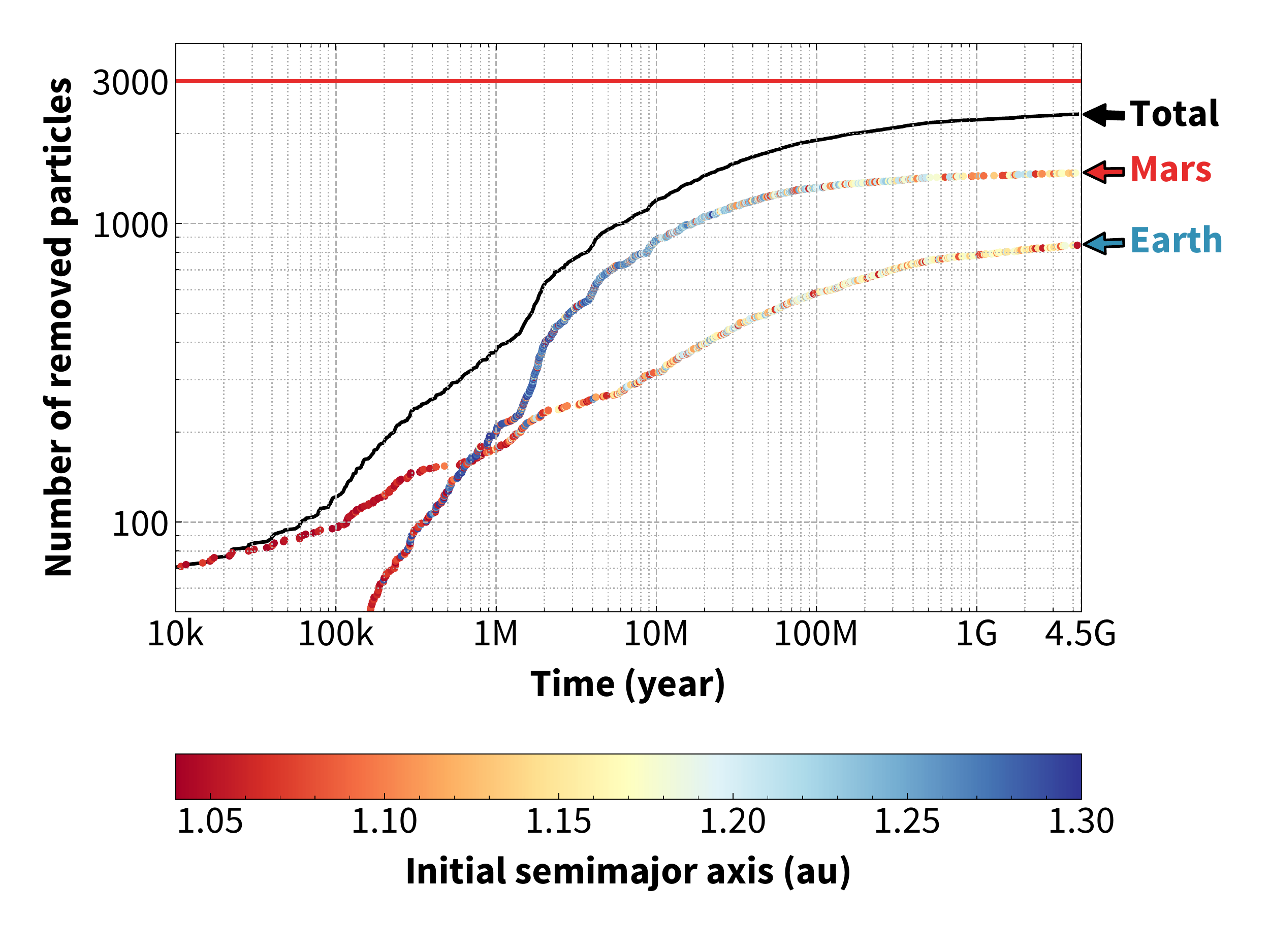}
	\caption{The removal history of test particles over 4.5 Gyr. The black solid curve gives the cumulative number of test particles removed as a function of time. The number of particles removed by Earth and Mars are separately shown via coloured dots, linked to their initial semimajor axes. The top red line represents the total number of integrated test particles.}
	\label{fig:discard_speed}
\end{figure}

Figure \ref{fig:discard_speed} provides the time history of particle removal.
%, plotted as the black curve, with removals by Earth and Mars separately shown in two different cumulative distributions.
Each test particle removal is colour-coded by its initial semimajor axis.
The early removal of particles in the vicinity of Earth (the resonance region) occurs before 1 Myr, which
indicates a rapid excitation from Earth scattering.
Between 10 Myr and 1 Gyr, most of the particles removed by Earth are coming
from the Earth--Mars belt; they need only
small $e$ increases (Fig.~\ref{fig:alltp}) to cross Earth's orbit.
By the simulation's end, only sporadic removals are occurring, again mostly from the Earth--Mars belt
region but equally shared between the two planets.

The unstable region beyond 1.17 au, on the other hand, tells another story. The vast majority of test particles from this region are removed
by Mars in the period of 1 Myr to 100 Myr, implying secular effects are likely involved in a long-term destabilization process.
\citet{2002MNRAS.333L...1E} suggested that two secular resonances, the $\nu_6$ and $\nu_{16}$, may lead to instability across the unstable region.
Their claim was based on a calculation of the location of linear secular resonances by
\citet{1997Icar..128..230M}, in which those authors show the presence of $\nu_6$ and $\nu_{16}$ resonances around 1.2 au for near-planar orbits with $e=0.1$.
However, no analysis of resonant angles was performed to explicitly identify those resonances.

Our main 4.5 Gyr integration used a 10,000~yr output interval which provided insufficient time sampling
for a detailed view of resonance arguments.
We thus examined the secular behaviour by carrying out another numerical simulation of 131 initially near-circular and planar orbits covering the unstable region from 1.17--1.30 au.
These orbits were integrated for 10 Myr with a denser output interval of 200 yr, enabling us to check both
the short-term and the long-term dynamical effects that could contribute to the instability.
Upon carefully examining the orbital history of these test particles, we find that not only
$\nu_6$, but also $\nu_3$ and $\nu_4$ help
drive up particle eccentricities,
causing orbital intersection and then encounters with either Earth or Mars.
The $\nu_3$ and $\nu_4$ resonances were located semi-analytically by \citet{1997Icar..128..230M} at
these semimajor axes, but at inclinations of $\simeq8^\circ$; our test particles rarely became inclined
to the Earth by more than 4 degrees.
Note that the third and the fourth Solar System eigenfrequencies  ($g_3$ and $g_4$) are
nearly equal and thus the $\nu_3$ and the $\nu_4$ locations are always close in orbital
parameter space.

Inspection of the resonant angles ({\it e.g.,} $\varpi - g_3 t $) showed non-circulation
at time intervals where eccentricities were rising.
This occurred at inclinations of only a few degrees in the unstable zone,
thus indicating that the semi-analytical method slightly miscalculates the resonant
locations.
This method assumed circular, coplanar planetary orbits and that particle $e$ and $i$ do not
change over a secular cycle; both these approximations are not strictly true and thus one
expects the resonance locations to not be perfectly predicted.
This explains why we uncover the role of $\nu_3$ and
$\nu_4$, which were not identified by \citet{2002MNRAS.333L...1E}.

In a more detailed examination, we noticed that the eccentricity rise for particles at the
inner edge of the unstable zone (closer to 1.17 au) have faster precession driven by Earth's
proximity which produces a match to the faster $g_6$ frequency.
We observed that particles with semimajor axes closer to 1.3~au were those whose
eccentricity rise was related to the slower $\nu_3$ and $\nu_4$.
This is reasonable, as a minor body further from a gravitational perturber tends to precess
slower.
The dominance of $\nu_6$ near 1.2~au and $\nu_3/\nu_4$ further out is clear in our
integrations, although the orbital evolutions are `messy' with many dynamical effects
operating simultaneously; for this reason we have simply labeled
Figure~\ref{fig:alltp}'s upper panel as being affected by `secular resonances' in
this unstable region.

Two clusters of surviving particles in the unstable region are seen in Figure \ref{fig:alltp}, at the
locations of the 7:9 and 3:4 mean-motion resonances with Earth.
However, upon careful examination of the relevant resonant arguments, we find that these test particles are not currently inside (librating in) these two mean motion resonances; the clusters are actually just beyond
resonance's borders.
%This is shown by two peaks and one dip around the 3:4 mean motion resonance in the third panel.
The orbital histories reveal these particles that are near mean-motion resonance are precessing
at (an expected) faster perihelion precession rate, resulting in them staying safely away from the highest eigenfrequency, $g_6$.
On the other hand, particles initially librating in these resonances
are found to become unstable over 4.5 Gyr.

Let us now focus on the main structure of the Earth--Mars belt (1.09 au $< a <$ 1.17 au).
Statistically speaking, near 60\% of test particles in this region survive for 4.5 Gyr,
leaving a metastable belt just exterior to the `resonant region'.
The two biggest features in the Earth-Mars belt are related to
major first-order mean-motion resonances (the venusian 1:2, the Earth's 4:5, and the martian 3:2) which, as \citet{2002MNRAS.333L...1E} previously pointed out, correspond to semimajor axes with unstable orbits.
The locations of the latter two are coincidentally close to each other, likely leading to overlap.
Resonance overlap is unlikely an important factor in the venusian 1:2 case;
this powerful resonance need only increase $e$
to $\simeq$ 0.1 to cause instability because the venusian resonance provides no phase
protection against Earth encounters.

The other first-order resonances in the belt (the 5:6, 6:7, and 7:8 with Earth), on the other hand,
do not appear to strongly destabilize those semimajor axes in the Earth--Mars belt.
Although a dip corresponding to the location of the 5:6 resonance is shown on the plot,
we think it is likely a statistical fluke, with only $\sim$10 test particles initially
present in that bin.
Hence, we conclude that these three resonances are not a major influence on
the belt.

\section{The nearby NEO Population}\label{sec:nea}

\begin{figure}
	\includegraphics[width=\columnwidth]{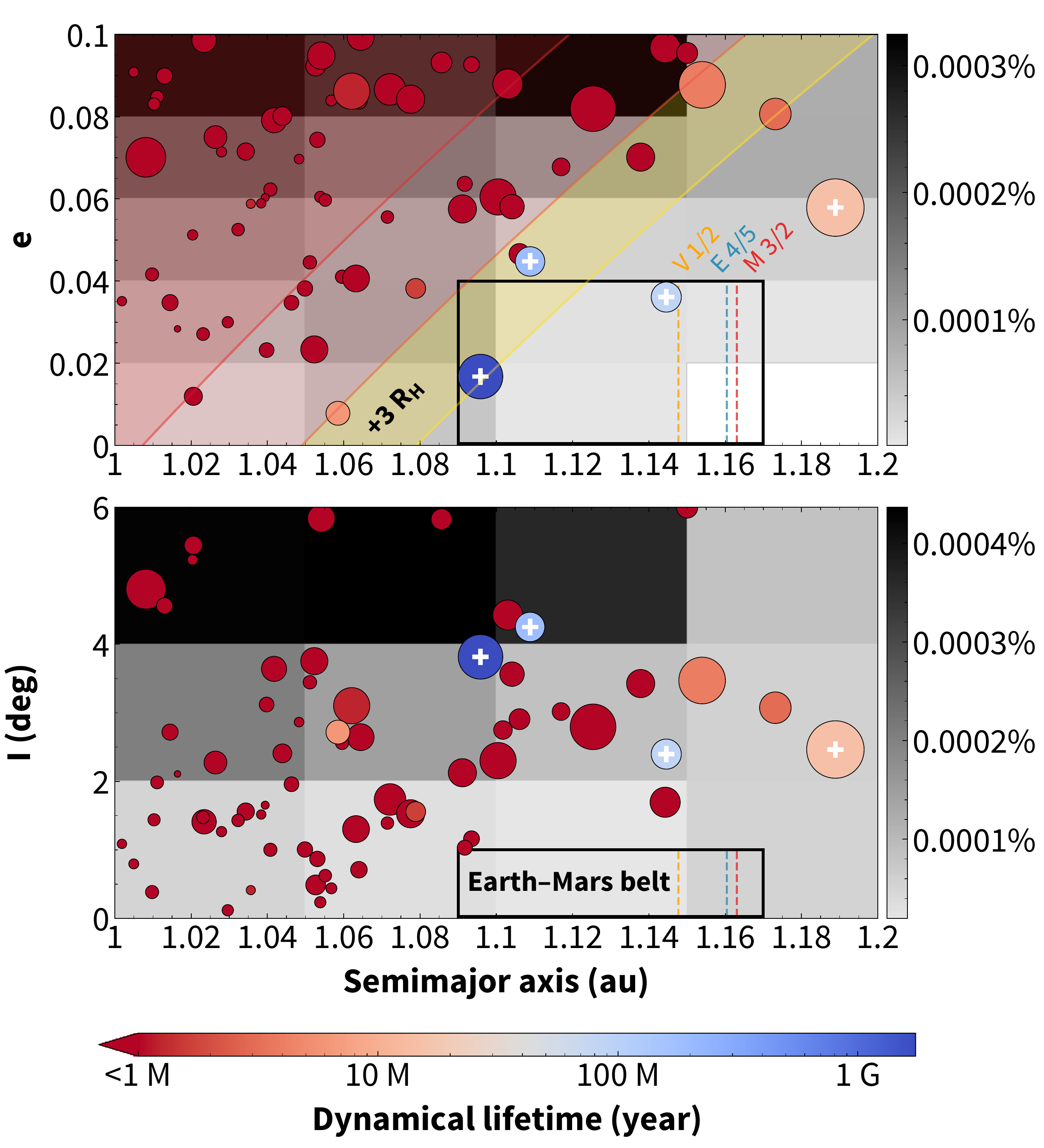}
	\caption{The $(a,e)$ and $(a,I)$ distributions of all NEOs surrounding our interested Earth--Mars belt.
		Each object is colour-coded by its dynamical lifetime and plotted with the diameter roughly estimated by its H magnitude;
		%To give the readers a sense of their relative sizes,
		the largest circle represents a $\sim$150 m asteroid and the smallest a $\sim$2 m one. White crosses mark four NEOs whose stability lifetime exceeds 10 Myr. Our boundaries of the stable belt are shown with rectangular outlines.
		Dashed vertical lines mark three important mean-motion resonances that affect belt stability (Fig.~\ref{fig:alltp}).
		The two upper-left red bands (same as in Fig.~\ref{fig:alltp}) denote Earth orbit intersection over Earth's $e$ variation, and the expected additional reach over +3 $R_H$  (yellow highlight) is where longer-lived metastable behavior is confirmed.
		The grayscale heatmap in the background shows projections of the residence time probability distribution from \citet{Greenstreet:2012ee} for where NEOs are expected;
		%at the coarse cell volumes of 0.05 au $\times$ 0.02 $\times$ $2\degree$;
		the greyscale represents the percentage of steady-state NEOs in each cell, with the white cell corresponding to zero (that is,
		there are no objects in this $(a,e)$ cell at any $I$).}
	\label{fig:asteroids}
\end{figure}

We have demonstrated that, considering only gravitational forces, a primordial Earth--Mars belt from $a=$1.09--1.17~au will retain $\approx$60\% of its population until the present day.
Thus, should any asteroids be identified in this $a$ range
with $e<0.04$ and $i<1^\circ$, they could in principle be
primordial.
We examined the current JPL Small-Body Database\footnote{JPL Small-Body Database (\url{https://ssd.jpl.nasa.gov/sbdb_query.cgi}), retrieved in November, 2019.},
and as of the time of writing, no minor planets exist
with orbits in this region.
However, there are minor planets close to this region,
and in this section we discuss the context of
NEOs near the Earth--Mars belt,
establishing that these NEOs are simply
members of the continuously-renewing Apollo and Amor
population in steady state.
Recall that in NEO orbital classification,
Apollos have $a>1$~au but $q<1.017$~au, while the Amors
have
%$a >$ 1.017 au and
1.017 au $< q <$ 1.3 au.

%However, it is possible that the belt region could be (temporarily) entered by objects initially exterior to the region and these objects would not be primordial.
%In this section, we show that no known NEOs currently exist in the Earth--Mars belt we defined in the last section (although a few are close);   and none of the nearby objects are likely to be a residual planetesimal population.

% A long-lived asteroid population must stay inside an orbital region that is long-term stable, or at least relatively stable compared to the age of the Solar System.
% However, the reverse is not necessarily true. The hypothetical long-lived belt does not ensure that real objects can be found. In this section, we will show that no NEOs actually exists in the Earth--Mars belt we defined in the last section, and none of the nearby objects are likely to be a residual planetesimal population.

%To verify whether or not any primordial asteroid still hides among the NEO population,

We retrieved all objects with
$a\sim(1,1.2)$, $e < 0.1$, $I < 6\degree$, and condition
code $\le 5$, from JPL Small-body Database\footnote{The eight objects with $a$=1.2--1.3~au that satisfy the same $e$ and $i$ restrictions are, in our view, NEOs temporarily dropped to small $e$ and $i$ by the secular resonances}, producing 66
asteroids.
Fig.~\ref{fig:asteroids} shows each NEO in $(a, e)$ and  $(a, I)$ phase space, using circles the radius of which corresponds to a roughly estimated diameter from its $H$ magnitude. The boundaries of the Earth--Mars belt on each panel are outlined with black rectangles. Despite two objects falling into the rectangle in the $(a, e)$ space, they are above the inclination range, and thus all known NEOs are outside the
stable Earth-Mars belt region that our integrations demonstrate.
(Note that our initial condition range was not chosen based on the NEO distribution, which was only examined after our integration.)

We directly integrated all these NEOs;
our results indicate that almost all
%surrounding the stable belt
have a very short stability lifetime\footnote{Here, we define the stability lifetime as the time until the object has a first close encounter with a planet.}, far less than 1 Myr.
In Figure \ref{fig:asteroids}, short-lived NEOs are colour-coded in red and almost all of them are unsurprisingly located inside the
Earth-intersecting zone (red bands in $a,e$ space).
Beyond this, one expects a metastable zone extending another
3 Hill spheres (3$R_H\simeq0.03$~au) where Earth's gravity
is expected to be able to eventually pull particles into close encounters.
The orbital stability of known NEOs from their forward integration
well matches these simple considerations.

Only four NEOs have dynamical lifetimes longer than 10 Myr and are marked by white crosses.
The asteroid 2011 AA$_{37}$ (dark blue circle) has the longest lifetime, of $\simeq$1.7 Gyr in our pure-gravity integration, and sits at the outer boundary
of Earth's $+3 R_H$ zone with a very low current $e\simeq$ 0.02 but
$I\simeq4^\circ$.
If there is any candidate for a primordial object it would be this object,
but with diameter of order 100~m, this asteroid's past and future dynamics are subject to Yarkovsky drift (see next
section).
The light blue circle in Fig.~\ref{fig:asteroids} near the venusian 1:2 resonance marks the location of 2019 AP$_8$.
Upon closer inspection of our integration, we find that it is this resonance that causes the eccentricity to increase and eventually leads to Earth crossing.
This corroborates the causality between the instability earlier detected at this semimajor axis (Fig.~\ref{fig:alltp}) and
this resonance.
Lastly, the rightmost cross is (225312) 1996 XB$_{27}$ and is
also the largest NEO known in this region.
Although currently far from Earth-crossing, it is situated
outside the Earth-Mars belt and is located in the `unstable
region' where our main integrations showed that secular resonances destabilize orbits within tens of million years.
Nothing in this analysis of real NEOs is thus in conflict
with our early conclusions in Section \ref{sec:numerical}.

Instead of looking at the future evolution of known NEOs,
a complementary viewpoint is provided by considering where the locations in
($a,e,I)$ space NEOs are expected to reach after exiting
the main asteroid belt.
The steady NEO flux is caused by the continuous removal of asteroids
from the main belt via resonances, delivering
them to planet-crossing orbits, some of which reach semimajor axes
near the Earth--Mars belt.
Previous studies \citep{2002Icar..156..399B, Greenstreet:2012ee, Granvik:2018jc}
have computed which portions of orbital parameter space are visited
by the NEOs during this dynamical evolution.

To illustrate where these models (which are not based on integration of
known real objects) predict NEOs would reach,
we used the NEOs orbital distribution model published by \citet{Greenstreet:2012ee},
which compiled the fraction of the NEO steady state population in $a,e,i$
cells of size 0.05 au $\times$ 0.02 $\times$ $2\degree$.
Figure \ref{fig:asteroids}'s background grayscale indicates the
summed percentage of the time steady-state NEOs spend in that portion of
phase space (projecting all $I$ values onto the $a,e$ plane and all $e$
values into the $a,I$ plane).
%In other words, the darker a cell is, the higher probability you have to find a NEO in that projected cell.
The known NEOs obviously all inhabit regions where the orbital model
shows have dynamics can deliver them to, making it completely plausible
that all these NEOs are recently delivered small main-belt asteroids,
the vast majority of which will remain only briefly ($<1$~Myr) near their
current orbit.
Unsurprisingly, the Earth-intersecting orbital element space is more heavily occupied
by this continuously moving population.
But NEOs can only reach the high-$a$ and low-$e$ portion of the parameter
space if resonances assist in reducing $e$, which is the inverse of the process
discussed above for the already low-$e$ real NEOs and likely how
these objects reached their current orbits.

This analysis supports the idea that the small Earth--Mars belt region
is not effectively reached during NEO evolution.
In fact, in Fig.~\ref{fig:asteroids}, the two cells that have $a$ = 1.10--1.15~au
and $e<0.04$ are only grey due to steady-state contributions
with $I>1^\circ$.
If we restrict to $I<1^\circ$ there are no steady-state NEOs in this $a,e$
range (like the bottom right cell which also has an estimated fraction of
zero).
Thus, the Earth--Mars belt is essentially unreachable by
main-belt visitors.

Because the known NEOs do not have a long dynamical lifetime, 
and because these NEO orbits are reachable from the main belt,
we think the 66 objects discussed here are very unlikely to be primordial.

\section{Yarkovsky Drift}\label{sec:discussions}

Given the extensiveness and depth of NEO surveys that would be
extremely sensitive to asteroids in the
$a=$ 1.09--1.17~au, $e<0.04$ and $I<1^\circ$ Earth--Mars belt
region,
there is no chance there is a substantial undiscovered population
of $>100$~m bodies.
In any case, even km-sized bodies would have their semimajor axes
drift substantially on Gyr time scales due to radiation pressure
effects.
How large would a primordial object need to be to prevent it from moving
the width of the belt over the age of the Solar System?

The Yarkovsky effect is a force exerted on a rotating minor body caused by the anisotropic emission of thermal photons.
It significantly alters the orbits of objects in the meter to ten-kilometer size range, moving them either inward or outward depending on the sense of rotation.
The significant role that Yarkovsky plays in meteoroid and NEO delivery
has been reviewed by \citet{Bottke:2006en}.
Slow Yarkovsky drift of semimajor axis delivers main-belt asteroids to
unstable resonance zones, which then deliver them to Earth-crossing orbits.
In a similar way, the Yarkovsky drift of primordial Earth--Mars belt asteroids
could have pushed them either
into the unstable gaps opened by mean-motion resonances
or beyond the belt's boundaries.
Therefore, the question we pose is: What is the smallest asteroid size that
will retain its stability for 4.5 Gyr under the Yarkovsky effect at the
Earth-Mars belt's location?

\citet{2020AJ....159...92G} reports detections of Yarkovsky drift for 247 NEOs,
which is the largest collection to date.
Their study covered a range of NEOs with various orbits, several of which are near
the Earth--Mars belt and can thus serve as references for our estimation.
In the assumption that the drift rate is unchanged during the entire evolution, we extrapolate their Yarkovsky drift rate measurements, in the unit of $\langle da/dt \rangle \sim 10^{-4}$ au/Myr, to give total drift $\Delta a$ over the age of
the Solar System.

\begin{figure}
	\includegraphics[width=\columnwidth]{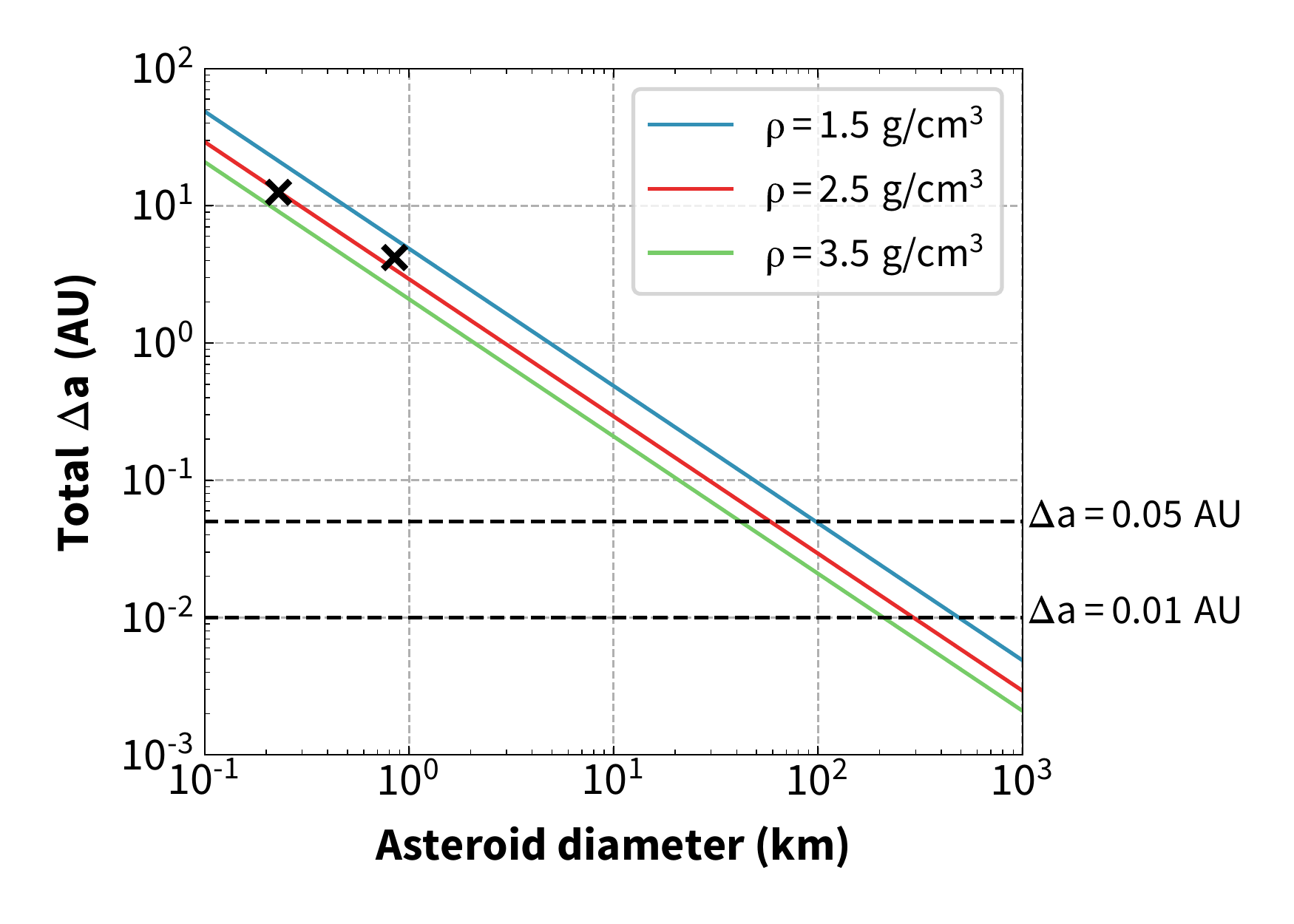}
	\caption{The total semimajor axis drift $\Delta a$ over 4.5 Gyr estimated for Earth--Mars belt asteroids of different sizes under the assumption $\xi = 0.12$. Three straight lines represent different asteroid densities and two black crosses mark the estimated extrapolated drift of the NEOs (474163) 1999 SO$_{5}$ and (452389) 2002 NW$_{16}$, studied in \citet{2020AJ....159...92G} and near the belt region. Two horizontal dashed lines show total drifts of 0.05 au and 0.01 au.
		Above $\Delta a = $ 0.05 au, almost all asteroids would eventually
		get removed by the drift, whereas below $\Delta a = $ 0.01 au, most asteroids
		would remain stable throughout the age of the Solar System.}
	\label{fig:yarkovsky}
\end{figure}

Re-scaling the Yarkovsky drift rate expression in \citet{2020AJ....159...92G}
results in the following equation for the 4.5 Gyr accumulated $\Delta a$ drift:
\begin{equation}
	\Delta a \bigg\rvert_\text{4.5} \! \! \! \! \! =
	\pm 0.065 \text{ au} \left( \frac{\xi}{0.1} \right)
	%	\left( 1 \text{ au} \over a\right)^{1\over2}
	\frac{1}{\sqrt{a_\text{au} }} \frac{1}{1-e^2} \left( \frac{100\text{ km}}{D} \right)
	\left( \frac{1 \text{ g/cc}}{\rho} \right)
\end{equation}
where $\xi$ is the Yarkovsky efficiency, $D$ and $\rho$ are the diameter and density of the object, respectively.
Figure \ref{fig:yarkovsky} shows the magnitude of the $\Delta a$ drift as a function
of asteroid diameter, where lines in different colour indicate different asteroid
density.
The calculation assumes $a = 1.13$~au, $e = 0$, and $\xi = 0.12$, where the latter is the median value of all NEOs studied in the report.
For reference, the measured Yarkovsky drift for two NEOs are converted to the total expected drift, and are shown in the figure as black crosses.
It should be noted that this is an over-simplification, as the
long-term Yarkovsky drift for small-sized bodies cannot be modeled by a constant drift rate, because obliquity, spin-rate, and $a$ changes all
alter the migration rate.

Unlike the main asteroid belt, which covers a broad orbital zone between Mars and Jupiter,
the region occupied by the hypothetical Earth--Mars belt is comparatively narrow,
with a width of merely 0.08 au (from 1.09 au to 1.17 au).
Moreover, two unstable gaps are present (Fig.~\ref{fig:alltp}), reducing
the maximum distance over which an asteroid could migrate while staying
in the belt down to 0.05 au (between 1.09 au to 1.14 au).
For this best-case scenario, Fig.~\ref{fig:yarkovsky} shows that $D>40$~km
is required (for a 3.5 g/cc object) for an asteroid to remain in the belt
over the age of the Solar System.
Slightly smaller asteroids with slower spin rates, particular obliquities,
or spin-pole reorientations due to collisions, could also survive.
Altogether, we confidently conclude that a lower size limit for a long-lived
Earth--Mars belt asteroid is $\sim 10$ km.
It is virtually certain that any object with $a\simeq1.1$~au
of this scale would already have been discovered by NEO surveys, and
thus we conclude that there are no primordial Earth-Mars belt objects.

%WE ARE HERE ---------------------------------
\section{Summary and Cosmogonic Implications}

In summary, our work has shown that the previously identified
\citep{2002MNRAS.333L...1E}
Earth--Mars belt is only stable over 4.5~Gyr time scale over
a restricted region: except for a small region from 1.04--1.09~au
where mean-motion resonances provide protection for a few objects,
the region (which we thus now re-define as the Earth--Mars belt)
with $e<0.04$, $I<1^\circ$ and $a=1.09-1.17$~au where about two thirds
of the particles in the simulations survive.
This region could thus hold on to large primordial objects over
the age of the Solar System, and is nearly unreachable by the
continuous flux of NEOs from the main belt.
We thus conclude that there is another stable belt in our Solar
System, but this stable belt is unoccupied today. For context, the roughly 10\% fractional width in semimajor axis
of the stable part of the Earth--Mars belt is comparable to the 4.5~au extent of the main Kuiper Belt at 45~au.

The fact that the stable Earth--Mars belt is devoid of large asteroids
provides some constraint on the end stage of terrestrial planet
formation.
Our result indicates that by the time terrestrial planets reached
their current orbits and masses,
at most a few $D>100$~km planetesimals could exist in the belt
region\footnote{More rigorously, $<5$ at 95\% confidence, for then
	if 60\% survive we would expect 3 and the Poisson probability of zero objects
	when the expected number is three is 5\%.}, or one should find
one today
(smaller objects can be removed by the Yarkovsky effect).

Because they focus on the final orbit distribution of fully formed
planets, studies that report on terrestrial planet simulations rarely publish the
($a,e,i$) distribution of the objects left over after planet
formation is largely complete.
What is clear from published models is that small bodies on circular coplanar orbits with $a$ = 1.1--1.2~au are
efficiently incorporated into the forming planets, and that
the rare survivors tend to have $e$ > 0.1
%\citep{ref:  }
% THIS ONE
%https://ui.adsabs.harvard.edu/abs/2011SSRv..163...41O/abstract
and thus also almost certainly $I>1^\circ$.
Thus, while current modeling makes it not surprising that no large
objects were in the Earth--Mars belt $\simeq4$~Gyr ago,
we suggest this can be used as an additional constraint on
the end states of such models.
In a broader perspective, this dearth of dynamically cold primordial
asteroids in the Earth--Mars belt connects well with the nearby lack
of stable Earth Trojan asteroids previously discussed, whose
primordial environment was similarly disruptive.

%ORDER OF MAG ESTIMATE of more that 100,000 objects in this a range
%if 100 km objects dominate the mass.   LET US NOT SAY THIS.

%Put a space station there!?     SHOULD WE USE THIS?:
Lastly, we point out that it would be possible to place a man-made
object at 1.1~au which would be dynamically stable for millions of
years with no need for station keeping.
This is the closest such highly-stable dynamical niche that would allow
repeated returns (at 0.1~au distance) to Earth, with a synodic period
of only a decade.

\section*{Acknowledgements}

We thank P.~Wiegert and S.~Greenstreet for valuable discusions, and an anonymous referee
for helpful improvements.
The authors acknowledge Canadian funding support from NSERC and China Scholarship Council grant number 201906210046.

\section*{Data availability}
Data available on request.
The data underlying this article will be shared on reasonable request to the corresponding author.

%%%%%%%%%%%%%%%%%%%%%%%%%%%%%%%%%%%%%%%%%%%%%%%%%%



\bibliographystyle{mnras}
\bibliography{embelt} % if your bibtex file is called example.bib


% Don't change these lines
\bsp	% typesetting comment
\label{lastpage}
\end{document}